\documentclass[11pt]{article}
\usepackage{blois,epsfig}

\def\be{\begin{equation}}
\def\ee{\end{equation}}
\def\bea{\begin{eqnarray}}
\def\eea{\end{eqnarray}}

\begin{document}
\vspace*{4cm}
\title{Nan\c cay radiotelescope as part of the international Pulsar Timing campaigns}

\author{I.Cognard, G.Theureau, G.Desvignes, R.Ferdman}

\address{LPC\small{2}E - CNRS/UMR 6115/OSUC, 3A, Av de la Recherche Scientifique\\
F45071 Orl\'eans CEDEX2, France}

\maketitle\abstracts{
Nancay radiotelescope is involved in high precision timing since 20 years. Since 2004, a coherent dedispersion
instrumentation enables numerous routine observations on more than 200 pulsars using half of the time if this
100-meters class radiotelescope. Two main programs are currently conducted.
A large set of young and old pulsars is timed for a multi-wavelength approach, complementary to the
very successful high energy observations of pulsars done by FERMI.
A set of highly stable millisecond pulsars is monitored as our contribution to the European Pulsar Timing Array
in order to probe the cosmological Gravitational Wave Background.
}

\section{Introduction}

Pulsars are highly magnetized neutron stars from which we receive collimated radio beams every rotation.
The most stable pulsars are used as clock for many different kind of studies.
Such pulsars located in a relativistic binary are used to put constrains on the Gravitation theories
\cite{2006Sci...314...97K}.
An array of highly stable pulsars distributed over the sky is used to search for a cosmological gravitational
waves background \cite{2006ApJ...653.1571J}.
An effort towards an international Pulsar Timing Array is taking place in every large telescope in
the world in order to share all the observations and set the best limit on
a gravitational waves background coming from the early Universe.

With a collecting area corresponding to a 100m dish, the Nan\c cay radiotelescope is a centimetric
telescope among the largest in the  world.
Based on a Kraus design, the telescope has two receivers providing a continuous coverage from 1.1 to 3.5GHz.
The 1.4 to 2GHz range is a good choice to observe pulsars between embarrasing interstellar effects
at the lower frequencies and faint emission received from pulsars at higher ones.

After a description of the Nan\c cay instrumentation, I will present few results in the context of
this search for a gravitational waves background.

\begin{figure}[t]
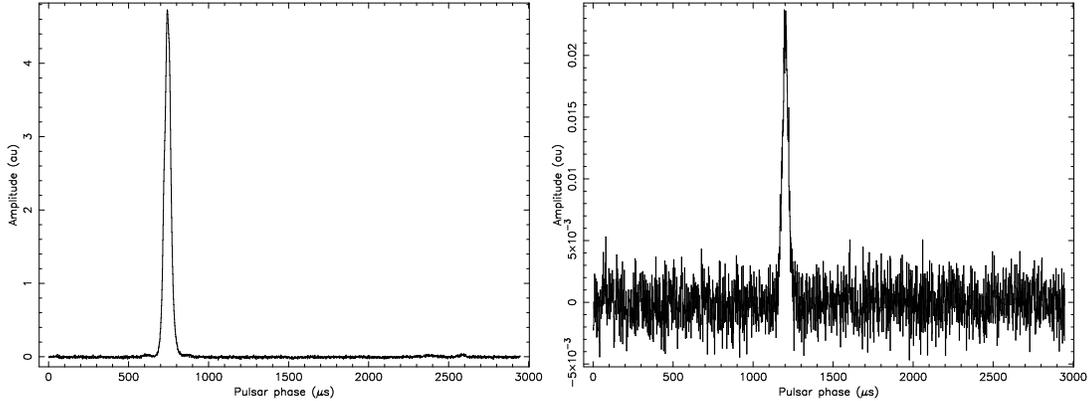

\begin{center}
\epsfig{figure=1909tpl.ps,angle=270,width=2.8in}
\epsfig{figure=1909-20090623.ps,angle=270,width=2.8in}
\caption{(left) Integrated profile obtained for the millisecond pulsar PSR J1909-3744 at Nan\c cay
used as a template to determine Times Of Arrival.
(right) A typical daily profile for the same pulsar, obtained on June 26, 2009.\label{fig:1909}}
\end{center}
\end{figure}

\section{Coherent pulsar dedispersion intrumentation}

A quasi-perfect way to remove the dispersion introduced by the ionized part of the interstellar medium,
called the coherent dedispersion,
is to apply a transfer inverse function in the complex Fourier domain of the data time serie
\cite{1975MComP..14...55H}.
This operation needs a huge computing power since we need to do direct and inverse
Fast Fourier Transforms in real-time on the Nyquist sampled data stream.
At Nan\c cay, we were the first to use the Graphical Processing Units (GPU)
instead of standard processors (CPU) to dedisperse pulsar data.
With the now fairly old Nvidia GeForce 8800GTX, we are able
to coherently dedisperse a 128MHz bandwidth having 2 computers hosting 2 GPU units each.
The two GPUs we put in each computer are water-cooled to increase their lifetime.
Presently the folding of the data in phase with the pulsar rotation is done in the CPUs.
and the overall load of each computer during observations is around 60\%.
Our present system is processing a bandwith of 128MHz, but we are already designing the
next generation which will be able to coherently dedisperse the 400MHz bandwith available
at the Nancay telescope.
While the current version is based on a Serendip5 board, we will
use a ROACH board also designed by the CASPER group (http://casper.berkeley.edu/).

\section{Multi-wavelength study}

A significant fraction of the pulsars observations done with the Nan\c cay radiotelscope
are about the young and noisy pulsars suspected to be detected at high energy by the
FERMI Large Area Telescope (LAT).
More than 100 pulsars are being regularly timed to produce, nearly automatically,
rotational ephemeris used to build $\gamma$-rays 'phasogram'.
The FERMI/LAT is very successfull with pulsars and 8 millisecond pulsars
were already detected \cite{2009Sci...325..848A}, using Nan\c cay ephemeris for 6 of them.
This is in addition to many younger and even radio-quiet pulsars which are also easily seen.

\section{an International effort}

The Nan\c cay radiotelescope is part of an european consortium called the European Pulsar Timing Array (EPTA)
which gather people from the five main radiotelescopes in Europe \cite{2008A&A...490..753J}:
Cagliari (IT), Effelsberg (G), Jodrell Bank(GB), Nan\c cay (F) and Westerbork (NL).
The 64 meters antenna in Sicilia (Cagliari) is still under construction.
We do have regular short workshops, usually at one of the radiotelescopes.
During a few days, we closely interact to make progress on the different actions we planned together :
coordinations of project and publications for students and post-docs,
coordinations of source lists and joined observing sessions,
sharing among partners of Time of Arrival (ToAs) and templates (used to derive ToAs),
building a library of common synthetic multi-frequency templates.
We are also open to international collaboration to share ToAs for joined goals,
and we are presently preparing a Memorandum of Understanding being signed with
the australian Parkes Pulsar Timing Array (PPTA). We hope to reach the same level
of agreement with the US NanoGrav consortium.

Lead by Michael Kramer (MPIfR, Bonn and University of Manchester), an advanced grant
from the European Research Council called LEAP (for Large European Array for Pulsars)
was obtained to coherently add the pulsar signal from the five large european radiotelescopes.
Based on the high concentration of large antennae in central Europe,
the aim is to get an Arecibo-like radiotelescope able to observe all the Northern sky.

\section{The Nan\c cay contribution}

The Nan\c cay pulsar coherent dedispersor is producing high quality timing data
on a number of stable millisecond pulsars.
Figure 1 shows an example of a template and a daily profile obtained on the ultra-stable pulsar PSR J1909-3744.
From the coherently dedispersed profiles,
we derive a precised ToA using a $\chi^2$ fit in the Fourier domain \cite{1992PTRSL.341..117T}.
A fit for the pulsar parameters (period and derivatives, position, proper motion and
potentialy several orbital parameters) is done from the ToAs using the code 'tempo2' \cite{2006MNRAS.369..655H}
producing ToA residuals (differences between calculated and measured ToAs).
We then carefully inspect them to make sure all the needed parameters are included in the analysis.
A set of ToAs residuals are shown Figure 2 with pulsars PSR J1600-3053, J1744-1134, J1857+0943, 1909-3744 and J1939+2134.
A summary of the ToAs residuals rms obtained at Nan\c cay are shown in Table 1 \cite{2009PhDDesvignes}.
The excellent quality of the Nan\c cay data can be emphasized observing that
half of the residuals are characterized by an rms below 1 $\mu$s and few of them are even below 500ns.
Nan\c cay is now an important partner to build a Pulsar Timing Array, within the EPTA first,
and then the EPTA being part of the International Pulsar Timing Array.

\begin{table}[t]
\begin{center}
\begin{tabular}{crrrcr}
\hline
Pulsar       & Period & P$_{orb}$  &    Span &   N$_{toa}$ &  $\sigma$   \\
             &  (ms) & (jours) &(yr) & & ($\mu$s)  \\
\hline
J0030$+$0451 &  4.87 &   $-$   &   4.6  &   402       &  1.84    \\
J0613$-$0200 &  3.06 &   1.2   &   4.5  &   280       &  0.913   \\
J0751$+$1807 &  3.48 &   0.26  &   4.5  &   158       &  1.73    \\
J0900$-$3144 & 11.10 &   18.7  &   2.0  &   199       &  2.87    \\
J1012$+$5397 &  5.25 &   0.6   &   4.3  &   107       &  0.771   \\
\\
J1022$+$1001 & 16.45 &   7.8   &   4.5  &   136       &  1.97    \\
J1024$-$0719 &  5.16 &   $-$   &   3.6  &   128       &  1.23    \\
J1455$-$3330 &  7.99 &   76.2  &   4.5  &   139       &  2.33    \\
J1600$-$3053 &  3.60 &   14.3  &   2.8  &   211       &  0.576   \\
J1643$-$1224 &  4.62 &   147   &   4.5  &   271       &  1.7     \\
\\
J1713$+$0747 &  4.57 &   67.8  &   4.5  &   260       &  0.350   \\
J1730$-$2304 &  8.12 &   $-$   &   4.5  &   85        &  1.55    \\
J1744$-$1134 &  4.07 &   $-$   &   4.5  &   87        &  0.343   \\
J1751$-$2857 &  3.91 &   110.7 &   3.5  &   36        &  0.948   \\
J1824$-$2452 &  3.05 &   $-$   &   4.5  &   313       &  2.63    \\
\\
J1857$+$0943 &  5.36 &   12.3  &   4.5  &   51        &  0.860   \\
J1909$-$3744 &  2.95 &   1.53  &   4.5  &   109       &  0.119   \\
J1910$+$1256 &  4.98 &   58.4  &   3.5  &   31        &  1.04    \\
J1939$+$2134 &  1.55 &   $-$   &   4.5  &   277       &  0.483   \\
J2145$-$0750 & 16.05 &   6.84  &   4.5  &   159       &  0.993   \\
\hline
\end{tabular}
\caption{Timing residuals rms for different pulsars monitored with the Nan\c cay radiotelescope.
Columns are pulsar's name, period, orbital period if binary, data span, number of ToAs
and ToA residuals rms.}
\end{center}
\end{table}

\begin{figure}[t]
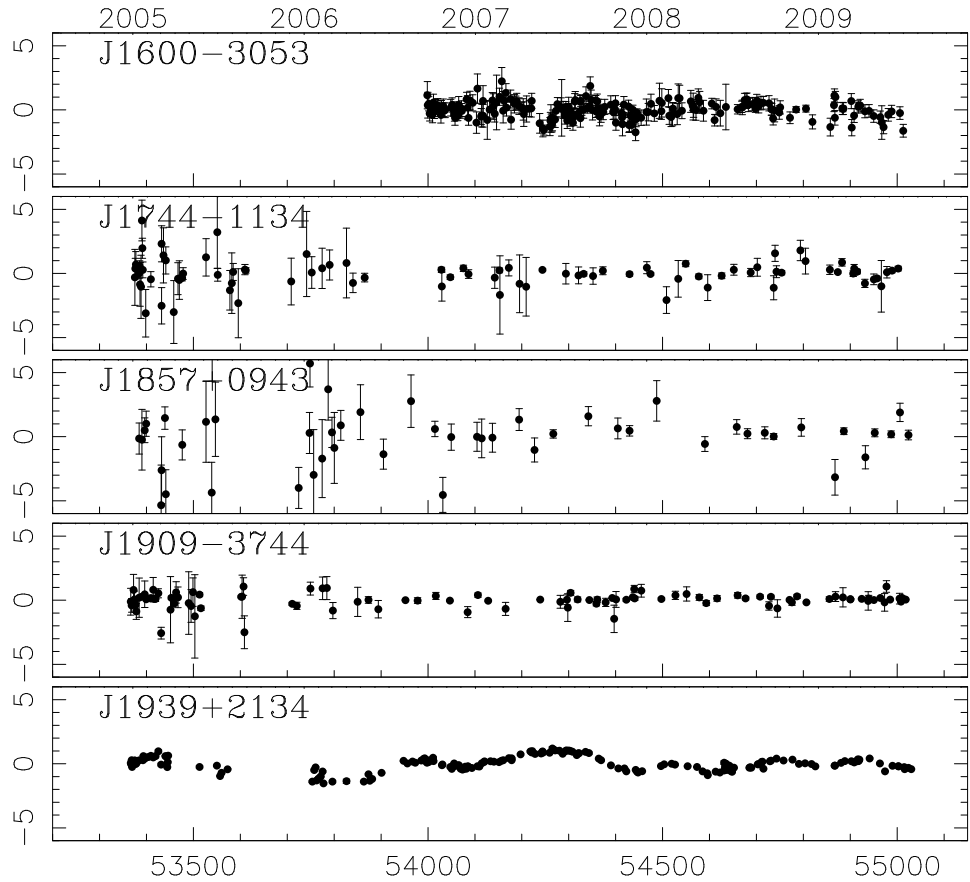

\begin{center}
\epsfig{figure=1600-b2009v.ps,width=5.0in}
\epsfig{figure=1744-b2009v.ps,width=5.0in}
\epsfig{figure=1855-b2009v.ps,width=5.0in}
\epsfig{figure=1909-b2009v.ps,width=5.0in}
\epsfig{figure=1937-b2009v.ps,width=5.0in}
\caption{Timing residuals for pulsars PSR J1600-3053, J1744-1134, J1857+0943, J1909-3744 and J1939+2134.
\label{fig:radish}}
\end{center}
\end{figure}

\section*{Acknowledgments}
The Nan\c cay Radio Observatory is operated by the
Paris Observatory, associated with the French Centre National de la
Recherche Scientifique(CNRS).

\end{document}